%% file: main.tex
\documentclass[conference]{IEEEtran}
\IEEEoverridecommandlockouts
\usepackage{svg}
\usepackage{cite}
\usepackage{amsthm}
\usepackage{amsmath,amssymb,amsfonts}
\usepackage{algorithmic}
\usepackage{graphicx}
\usepackage{textcomp}
\usepackage{xcolor}
\usepackage{bm}
\usepackage{caption}
\usepackage{subcaption}
\usepackage{multirow}
\usepackage{multicol}
\usepackage{caption}
\usepackage{subcaption}
\usepackage{geometry}
\begin{document}

\title{\Large\bf Analog/Mixed-Signal Circuit Synthesis Enabled by the Advancements of Circuit Architectures and Machine Learning Algorithms}


\author{\IEEEauthorblockN{Shiyu Su, Qiaochu Zhang, Mohsen Hassanpourghadi, Juzheng Liu, Rezwan A Rasul, and Mike Shuo-Wei Chen}
\IEEEauthorblockA{\textit{Ming Hsieh Department of Electrical and Computer Engineering}\\
\textit{University of Southern California}, 
Los Angeles, CA 90089 \\
\{shiyusu, qiaochuz, mhassanp, juzhengl, rrasul, swchen\}@usc.edu}
}


\maketitle




{\small\bf Abstract---
Analog mixed-signal (AMS) circuit architecture has evolved towards more digital friendly due to technology scaling and demand for higher flexibility/reconfigurability. Meanwhile, the design complexity and cost of AMS circuits has substantially increased due to the necessity of optimizing the circuit sizing, layout, and verification of a complex AMS circuit. On the other hand, machine learning (ML) algorithms have been under exponential growth over the past decade and actively exploited by the electronic design automation (EDA) community. This paper will identify the opportunities and challenges brought about by this trend and overview several emerging AMS design methodologies that are enabled by the recent evolution of AMS circuit architectures and machine learning algorithms. Specifically, we will focus on using neural-network-based surrogate models to expedite the circuit design parameter search and layout iterations. Lastly, we will demonstrate the rapid synthesis of several AMS circuit examples from specification to silicon prototype, with significantly reduced human intervention.}



\IEEEpeerreviewmaketitle

\section{Introduction}

\begin{figure}[!t]
    \centering
    \includegraphics[width=0.38\textwidth]{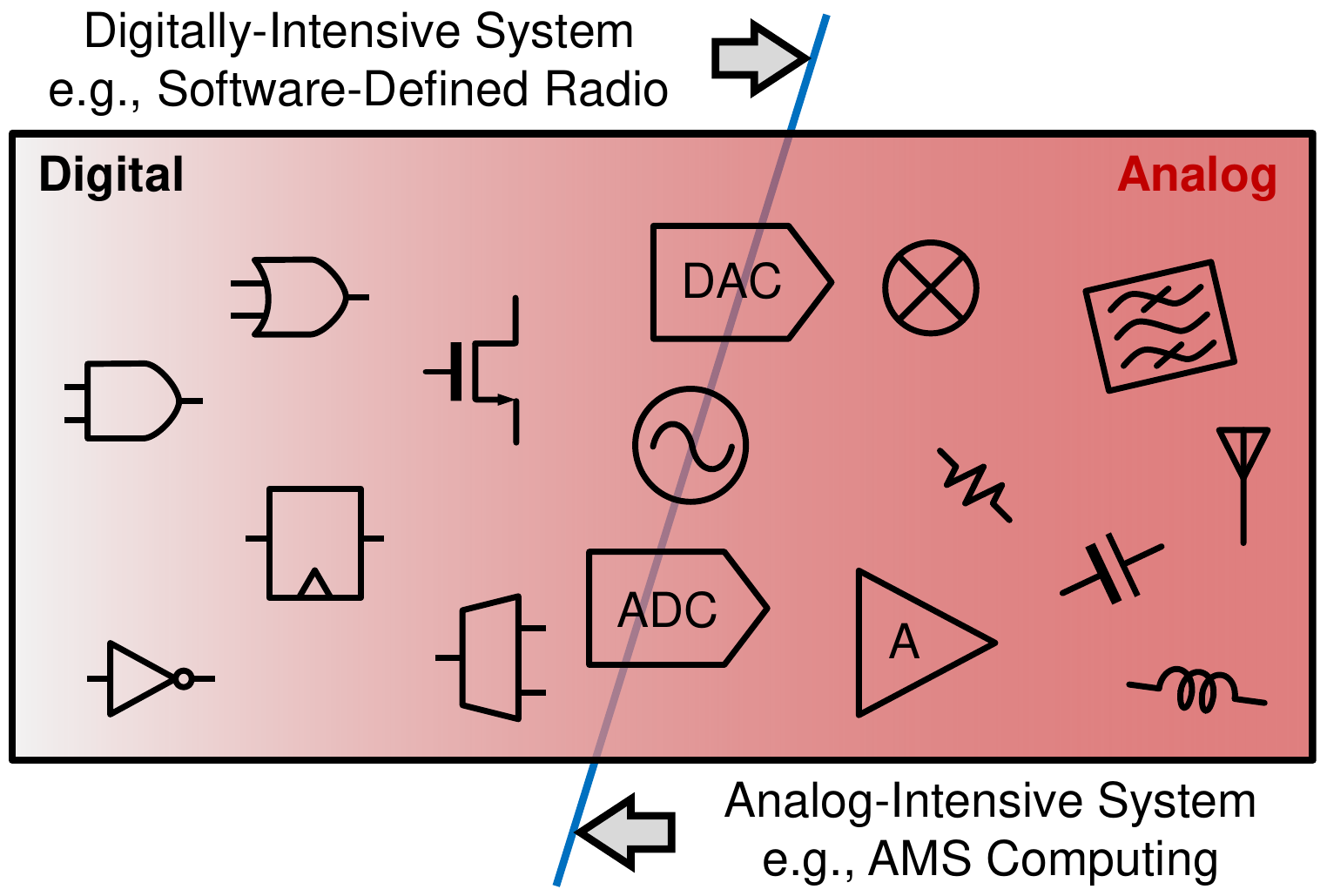}
    \caption{Blurring interface of System-on-Chip (SoC) design.}
    \label{blurring_interface}
\end{figure}

In traditional circuit design, there are clear boundaries between the digital back-end, analog mixed-signal (AMS) and radio frequency (RF) front-end circuits. As we are approaching the limits of CMOS technology scaling in terms of device size and power efficiency, improving the performance of conventional AMS circuits becomes incredibly challenging and inefficient. Therefore, circuit designers resort to architectural and/or system-level re-thinking. Consequently, co-design and co-optimization across devices, circuits, and algorithm have spawned significant number of innovations in interfaces (i.e., AMS) design. Driven by the growing performance and efficiency requirement of communication and computing system, the boundaries between analog and digital domain are blurring (Fig.~\ref{blurring_interface}). As a result, AMS circuits, especially data converters, become crucial to various emerging systems that need to cross between analog and digital domains. In a nutshell, the industry demands AMS circuits across a wide specification range (i.e., performance, power and area). However, the high degrees of freedom for optimizing such circuits poses a great challenge to deliver optimized designs within a reasonable time frame. In addition, the increasing design cost in advanced technology nodes further necessitates the reduction of time to market \cite{kahng2018new}, motivating AMS circuit synthesis.

However, the complexity for AMS circuit synthesis is generally higher than digital circuit synthesis. For example, constrained by the accuracy requirement of both continuous amplitude and time, simulations of analog circuits take significantly longer than that of digital circuits. Moreover, since the primitive design unit is down to transistor level instead of discrete digital standard cells, the parameter space of analog circuits is enormous compared to its digital counterpart, which demands substantially more iterations to achieve an optimum design. In addition, AMS design phases, including behavior modeling, schematic design and layout, require close guidance by the analog circuit designers, further increasing the design time. All those factors set a higher barrier for AMS circuit synthesis.

On the other hand, AMS circuit has gradually moved towards digital-intensive, analog-lite architectures to leverage the benefits of technology scaling maximally and achieve high flexibility and enhanced performance simultaneously. The digital-intensive AMS circuit architectures enable the possibility of leveraging digital design flow to synthesize complex AMS circuits like data converters, phase-locked loops, and digital transceivers. Meanwhile, the advancement of machine learning (ML) algorithms has been exploding over the past decade. Many algorithmic innovations have resulted in significantly improved accuracy for various modeling and classification tasks.

\begin{figure*}[!t]
    \centering
    \includegraphics[width=0.8\textwidth]{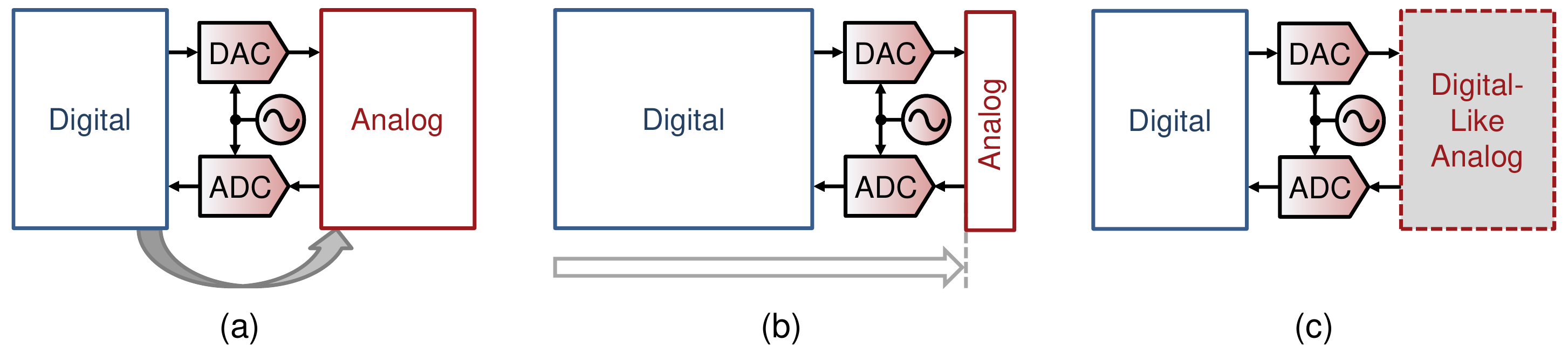}
    \caption{(a) Digitally-assisted (b) mostly digital and (c) digital-like architectures.}
    \label{digital_architectures}
\end{figure*}

This shift in AMS circuit architecture along with the recent advances in ML algorithms has provided a new opportunity for AMS circuit synthesis with high dimensional optimization, despite the aforementioned difficulties for AMS design. Moreover, a recent move toward open-source circuit design, including EDA tools and IPs, can potentially facilitate AMS circuit synthesis.  In this paper, we will broadly review the emerging architectures and ML algorithms suitable for AMS circuit synthesis. Several representative synthesis examples will be provided. Lastly, we will discuss a potential open-source design ecosystem enabled by AMS circuit synthesis.


The rest of this paper is organized as follows. Section II reviews the mostly digital AMS architectures and the associated techniques that favor design automation, especially standard digital design tools and flows. Next section III discusses the new opportunities in rapid AMS circuit synthesis enabled by the deep learning algorithms, focusing on the NN-based surrogate model for circuit parameter search. Design examples are provided in section IV. Section V describes the vision on open-source AMS design, followed by section VI which concludes the paper.





\section{Digital-Empowered AMS Architectures}

The key motivation of pushing AMS circuits towards more digitally-intensive architecture stems from the fact that analog circuits cannot leverage the CMOS technology scaling intrinsically as much as the digital circuits, in terms of both circuit performance and design cost. Due to the limited benefits offered by the scaling, architecture innovation has been the main driver of AMS circuit/system performance improvement \cite{manganaro2018emerging} and \cite{robertson2016data}. As the CMOS technology has advanced to 5nm and below, the short-channel transistors continue to favor mostly digital AMS architectures with performance and cost advantages \cite{loke2018analog}.


To illustrate the recent evolution of AMS circuits, we roughly divide the AMS architectures into three categories, as shown in Fig.~\ref{digital_architectures}. Starting around year 2000, applying digital signal processing techniques to assist or relax the analog circuit design became an active area of research (Fig.~\ref{digital_architectures}(a)). Motivated by \cite{mitola1995software} circuit designers pushed the performance of data converters and clock, which aimed to replace most analog signal conditioning by digital signal processing (DSP), making the system highly flexible (Fig.~\ref{digital_architectures}(b)). However, depending on the application, extremely high-performance data converters and PLLs might diminish the overall system efficiency. In such scenario, keeping some analog conditioning in the system while approximating the analog behaviors with digital-like operations can be a promising alternative (Fig.~\ref{digital_architectures}(c)). In the rest of this section, we elaborate those three types of AMS architectures in the context of AMS circuit synthesis.

\subsection{Digitally-Assisted AMS Design}
A major challenge in an AMS design is the fundamental trade-off between the area of the device and its mismatch. Larger device provides better matching but also leads to higher cost and lower speed. As transistors have been scaled down to 65nm and smaller, digital signal processing can relax the matching requirement of analog circuits with decent power- and area-efficiency. In \cite{murmann200312}, digital calibration is used to relax the precision requirement of the residue amplifier in a pipeline analog-to-digital converter (ADC) for significant power and area saving. Likewise, \cite{chiu2004least} proposed a background calibration technique based on adaptive filters to compensate for the nonlinearity of analog circuits in the ADC. More comprehensive calibration techniques have enabled new regime of high-performance ADCs \cite{ali202012}. Similarly, advanced digital pre-distortion and noise shaping techniques have been developed for wideband and high dynamic range digital-to-analog converters (DACs) \cite{su201612,lin201816b,su201816}. In addition to the performance enhancement, the above digital calibrations also reduce the analog complexity and ease the design automation.



\subsection{Mostly Digital AMS Architectures}
\begin{figure}[!t]
    \centering
    \includegraphics[width=0.49\textwidth]{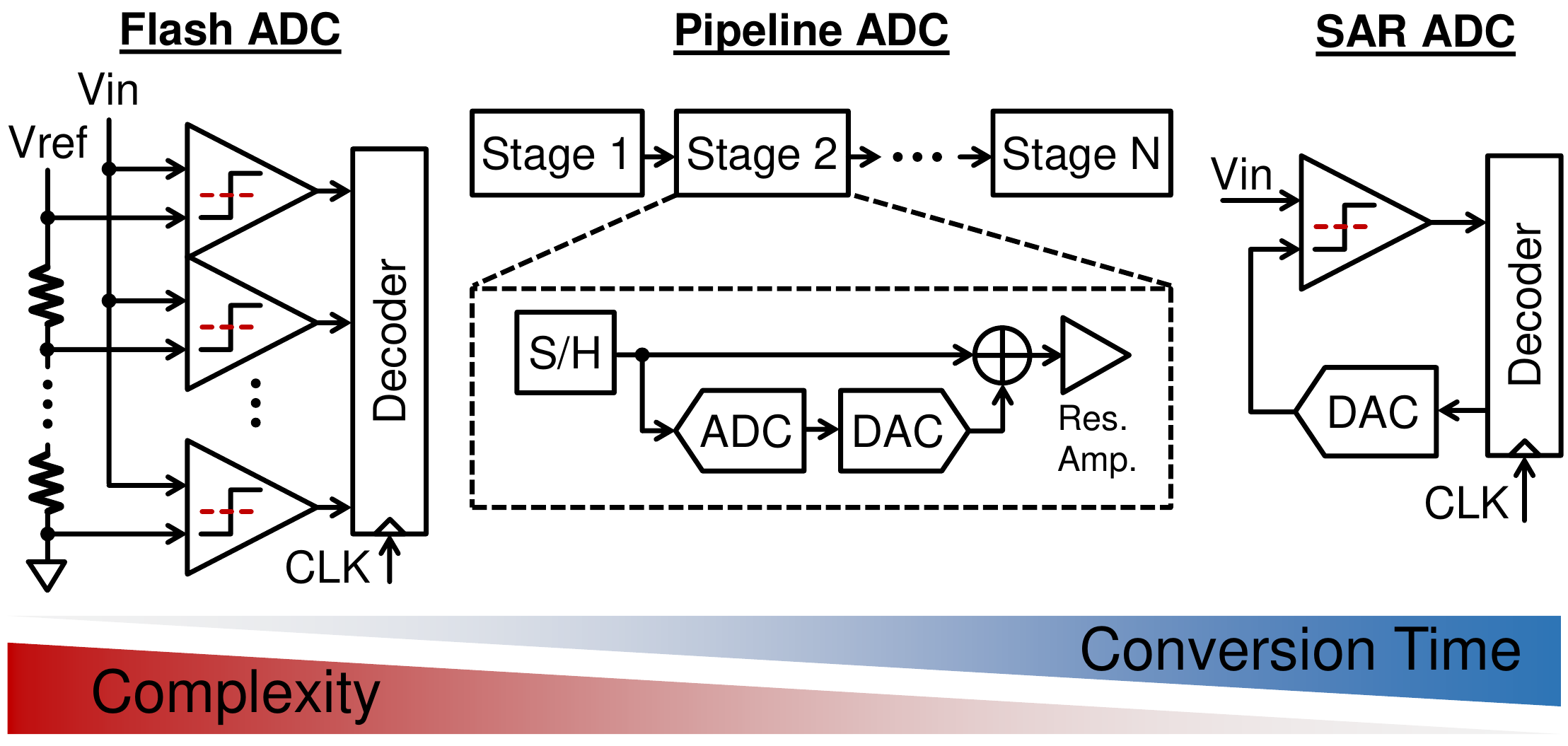}
    \caption{ADC architecture trade-off.}
    \label{adc_architectures}
\end{figure}

In parallel, designers have demonstrated mostly-digital architectures in the direct sampling receiver \cite{wu20162} and DAC-based transmitter \cite{spiridon2013375} using high-performance data converters for superior system flexibility. Such architectures have also been broadly explored for various AMS component blocks to leverage the increasing digital signal processing capability in advanced nodes. One such example is the digital phase-locked loop (DPLL), which has attracted much attention lately \cite{staszewski2005all,deng2014fully,ho2016fractional, chen2010calibration, ho2016digital, zhang2021fractional}. By pushing the control processing unit into digital domain completely, DPLL shows impressive robustness against process, voltage and temperature (PVT) variations and intrinsically allows digital calibration algorithms to improve the performance. More importantly, DPLL can be synthesized using standard digital design flow thanks to its mostly digital architecture. Fully synthesized DPLLs have demonstrated a significantly reduced implementation overhead with performance close to that of analog PLLs \cite{kundu2020self,deng2014fully}. Similarly, digital low-dropout regulator (DLDO) was proposed for low-noise and low-supply voltage applications \cite{okuma2010cicc}. Digitally-intensive dual-rate hybrid DAC was used to achieve high-speed and high-resolution simultaneously \cite{su201512}. Likewise, thanks to its minimum analog complexity among the ADC architectures, successive approximation register (SAR) topology has been widely adopted \cite{chen20066, ding2018circuit}. Since SAR ADC performs the conversion sequentially, the conversion rate inevitably slows down, as shown in the speed and complexity trade-off in Fig.~\ref{adc_architectures}. Time interleaving technique is typically utilized \cite{chang2018bag2} to boost up the rate.


\subsection{Digital-like AMS Operations} 

Another ongoing trend in AMS design is to use digital gates to achieve or approximate the analog functionalities in order to advance the circuit performance and reduce the design cost. Consider time-based ADC as an example. In recent years, the trend to operate time-based ADC above GHz sample rate has increased significantly. The ADC usually consists of a voltage-to-time converter for encoding the voltage information into time domain and a time-to-digital converter (TDC) for quantizing the time. The TDC is either a delay-line or a voltage-controlled oscillator \cite{dudek2000high, mohsen2018vcoadc, jssc_vcobased5g_2019,jssc_rnstadc_2018} and can be implemented by inverters and flip-flops only. Due to the smaller size of the digital circuits, fewer routing parasitics are expected in time-based ADCs. Moreover, digital circuits can achieve fast speed in advanced technology nodes without consuming too much power. As a result, the delay line based TDCs in \cite{mohsen2019ppstdc} and \cite{Juzheng2022Timebase} have reached up to 5GS/s using a single channel, which was previously only possible using Flash ADC or excessive paralleling (i.e., time interleaving), incurring significant area and power overhead. Along the same line, a design automation flow for a mostly digital voltage-controlled oscillator (VCO)-based delta-sigma ADC has been proposed and demonstrated recently \cite{xu2017scaling}. Custom library and flow were combined with the digital design flow and scaling benefits were shown by comparing different processes. Likewise, \cite{chen2019digital} proposed a complete design automation flow including logic synthesis, placement, and routing schemes for time-domain computing circuits. In similar manner, \cite{deng2014fully} utilized NAND gates to implement the current digital-to-analog converter (DAC) for a current-controlled ring oscillator. A digital-based operational amplifier \cite{crovetti2013digital} was proposed as well, blurring the boundary between analog and digital circuits. Furthermore, a synthesized switched-R-MOSFET-C analog filter was demonstrated using digital standard cells \cite{Liu2015cicc}. In addition to these baseband circuit blocks, \cite{su2020jssc} and \cite{su2020taf} approximated the amplitude-varying (i.e., analog) impulse response of an RF filter with a constant amplitude but a time-varying binary (i.e., digital-like) impulse response, such that the frequency responses are similar within a certain band of interest. Based on the specifications, the impulse response of a target filter is first designed using standard digital filter design flows, such as the FDA tool in MATLAB, followed by the time approximation via pulse-width modulation (PWM). In principle, such digital-like or time approximated AMS circuits favor the digital EDA tools \cite{wei2021analog}, however specialized algorithms may be needed \cite{chen2019digital,su2022tafa}.



\section{NN-Assisted AMS Design}



\begin{figure}[!t]
    \centering
    \includegraphics[width=0.5\textwidth]{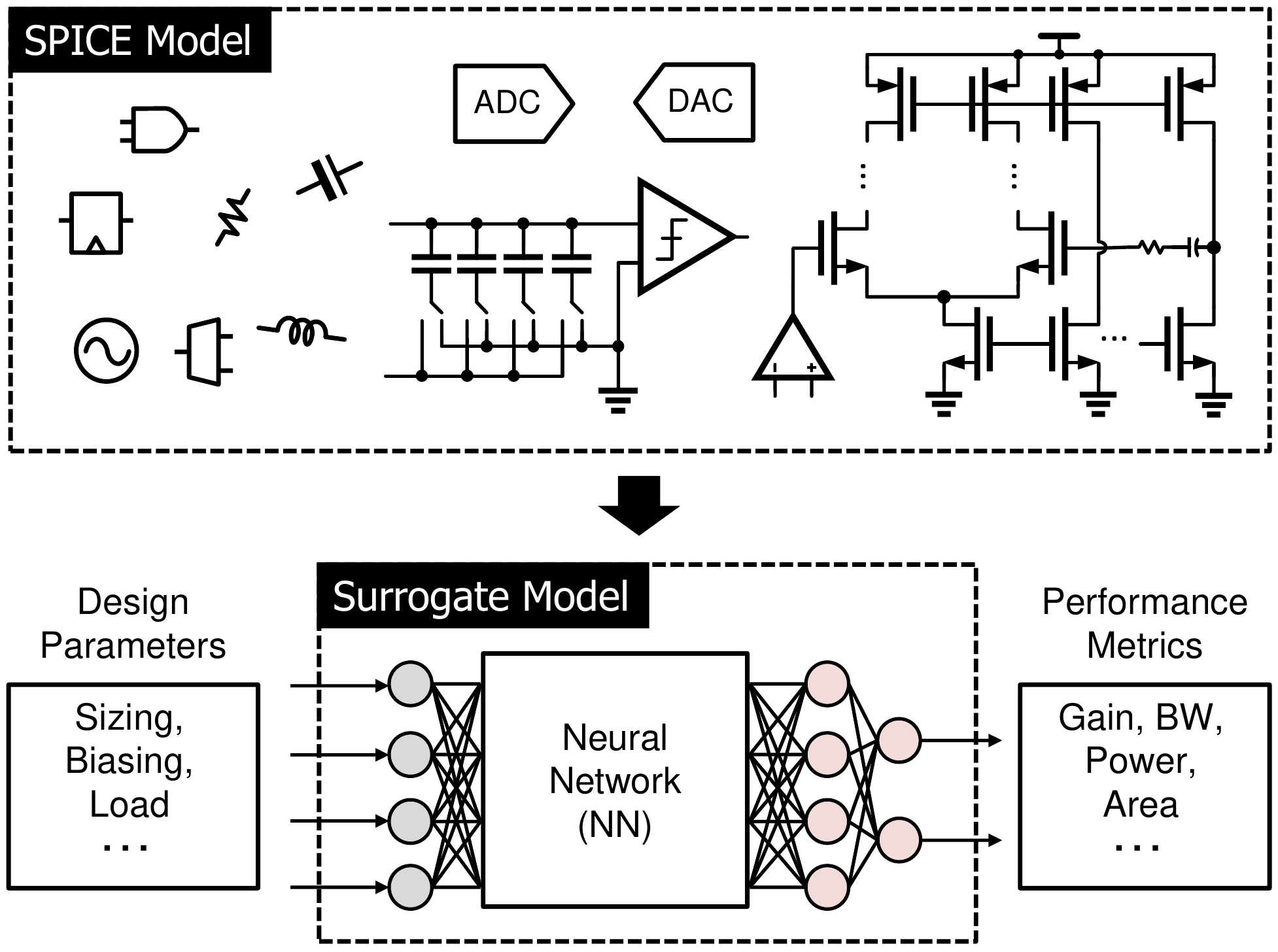}
    \caption{NN-based surrogate model.}
    \label{surrogate_model}
\end{figure}

To achieve a complete AMS circuit synthesis, one cannot solely rely on the architecture innovation by incorporating mostly digital design. New design methodology for AMS circuits is essential to tackle the grand challenges posed by advanced technology nodes (16nm and below), which results from the following observations: 
\begin{itemize}
    \item The device dimension is more discrete, yielding less degree of freedom for circuit sizing. 
    \item Layout design rule is more complicated and constrained and hence harder for manual design.
    \item The device model and the layout parasitic extraction are more complex, dramatically increasing the simulation time.
\end{itemize}
Consequently, it is extremely costly to design a close-to optimal AMS circuit. Therefore, AMS circuit synthesis with reduced design efforts and sufficiently good performance is highly desirable.

AMS synthesis cast a long research history with various approaches demonstrated in the past decades. The paper focuses the model-based methods due to their fast evaluation speed, reusability, and low computational cost. In the early days, the designers coded all the circuit knowledge in a hierarchical fashion\cite{harjani1989oasys} and synthesized relatively small circuit blocks like amplifiers. Geometric programming was also proposed to cast the Op-Amp design into a convex optimization problem \cite{boyd2001optimal} and later utilized for automating the design of analog PLL \cite{colleran2003optimization} and pipeline ADC \cite{del2002design}. Other surrogate models such as support vector regression \cite{de2005mixed,kiely2004performance}, neural network (NN) \cite{liu2002remembrance}, and Gausian process model \cite{wang2015bayesian,lyu2017efficient} have been widely explored for reducing the computational costs and model preparation overhead. Among the approaches, the NN regression outperforms others since it has more tunable hyperparameters, enabling accurate modeling of circuits which exercise a sophisticated non-linear function \cite{gielen2009tcad,li2020tcad1}. Therefore, NN has been deployed in many computer-aided design (CAD) tools. In the rest of this section, we elaborate on the use of NN-based surrogate model for AMS design.








\subsection{NN-based Surrogate Model and Parameter Search}\label{AA}

\begin{figure*}[!t]
    \centering
    \includegraphics[width=0.85\textwidth]{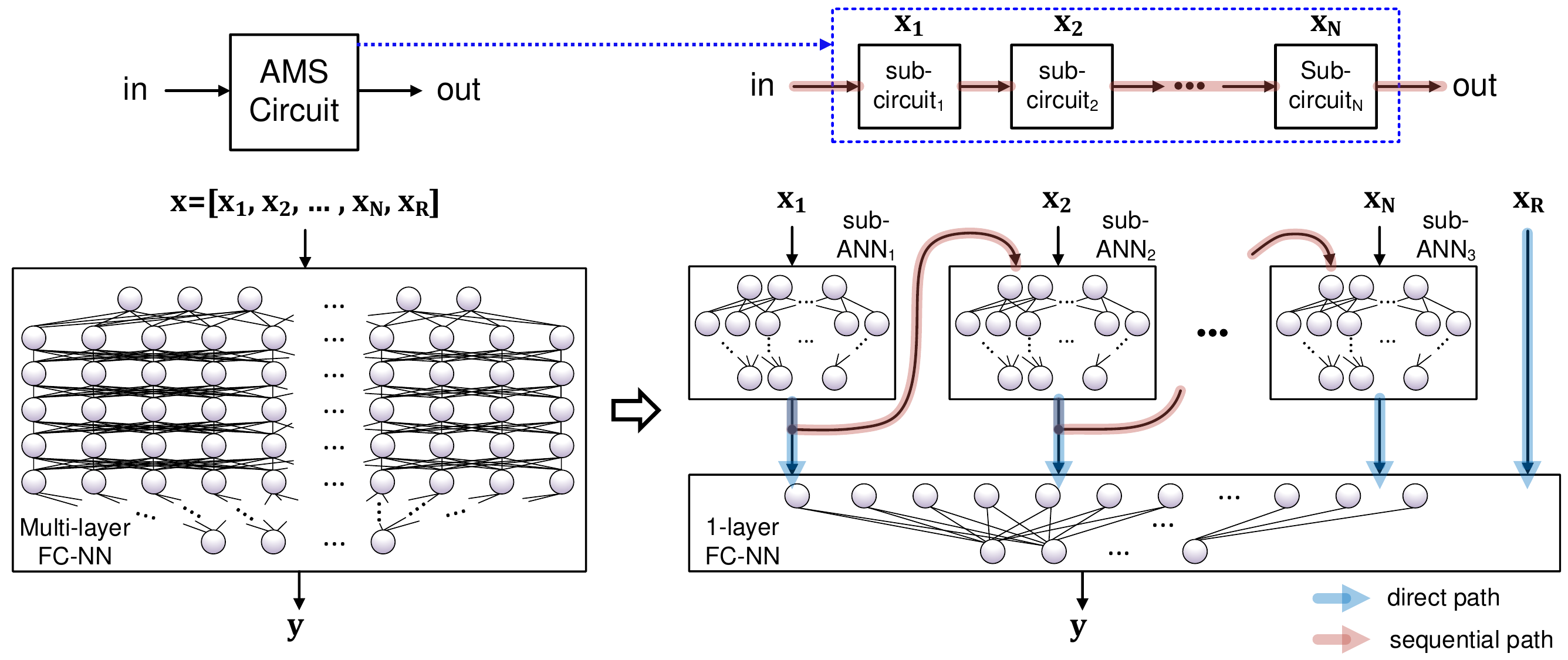}
    \caption{Concept of CCI-NN.}
    \label{CCINN}
\end{figure*}

 A surrogate model can replace the SPICE model to avoid expensive SPICE simulations, especially the post-layout simulations in advanced technology nodes (Fig.~\ref{surrogate_model}). A NN surrogate model was proposed to characterize the circuit's metrics in \cite{wolfe2003tcad1,islam2019icsmas}. A single NN model was used to predict the metrics of a circuit as simple as a single-stage amplifier or as complex as a PLL \cite{garit2012icvd}. Unfortunately, similar to other regression methods, NN exhibits an increase in regression error when the target circuit is larger. Therefore, the hierarchical design method divides a complex system into smaller sub-circuits called modules and models these modules using regression. With behavioral or functional models, the modules' metrics are then related to the system specifications. NN herein plays the role of module-level characterization \cite{modelbased_multiObj_2005}. In contrary, \cite{lorenco2019icsmas} used the NN to model the metrics-to-parameters function of the modules and used the trained model to set the initial parameter values for further optimization using SPICE simulations. \cite{li2020tcad1} suggested to perform a global search with the genetic algorithm using SPICE simulations at first, then train the NN model using data points in the vicinity of global search outcome, and finally perform local optimization using the trained model to further improve the performance. Although, the approach is efficient in enhancing the optimization speed, the NN model needs to be trained every time the global optimization is performed and cannot be reused as a result. 
 

In general, conventional hierarchical design fails to model the system properly when interactions among the modules become more extensive. Precise system modeling requires proper interface characterization, without which interface problem occurs. The module linking graph (MLG) concept first introduced in \cite{AMPSE_core_2021} accommodates a platform where the modules' interface can be part of the system modeling. MLG is a directed graph containing the modules as the vertices and the direction of the edges shows the cause and effect relations between two modules. Since estimating system specification with MLG requires many iterations, NN modeled modules are used in \cite{AMPSE_core_2021} but only for global optimization. Combining global optimization and sufficiently accurate NN models accelerates the search process while delivering nearly optimal results. After global optimization, \cite{AMPSE_core_2021} proposed to perform local optimization with SPICE simulations, removing the least significant parameters based on their gradients. The algorithm, called MOHSENN, can rapidly synthesize various AMS circuits with comparable or even better performance than manual design from an experienced designer.

The idea of MLG was further explored in \cite{mohsen2021acmdac}. Referred to as circuit connectivity inspired NN (CCI-NN), the NN structure was customized according to the circuit connection. The method achieves higher accuracy compared to the conventional fully-connected network (Fig.~\ref{CCINN}) given the same number of training data. Alternately, CCI-NN requires less training data to achieve the same model accuracy as a fully-connected network. Also, the network only requires a single dataset generated from the system simulations and does not need multiple training dataset for the modules and behavioral or functional modeling between modules' metrics and systems' specifications. CCI-NN can inherently learn the module-to-system relations and model the interfaces among the modules better. \cite{mohsen2021acmdac} showed that for proper modeling of an 8-bit 20GS/s current-steering DAC, CCI-NN required at least four times less training data compared to the regression models using conventional fully-connected NN.



\begin{figure}[!t]
    \centering
    \includegraphics[width=0.42\textwidth]{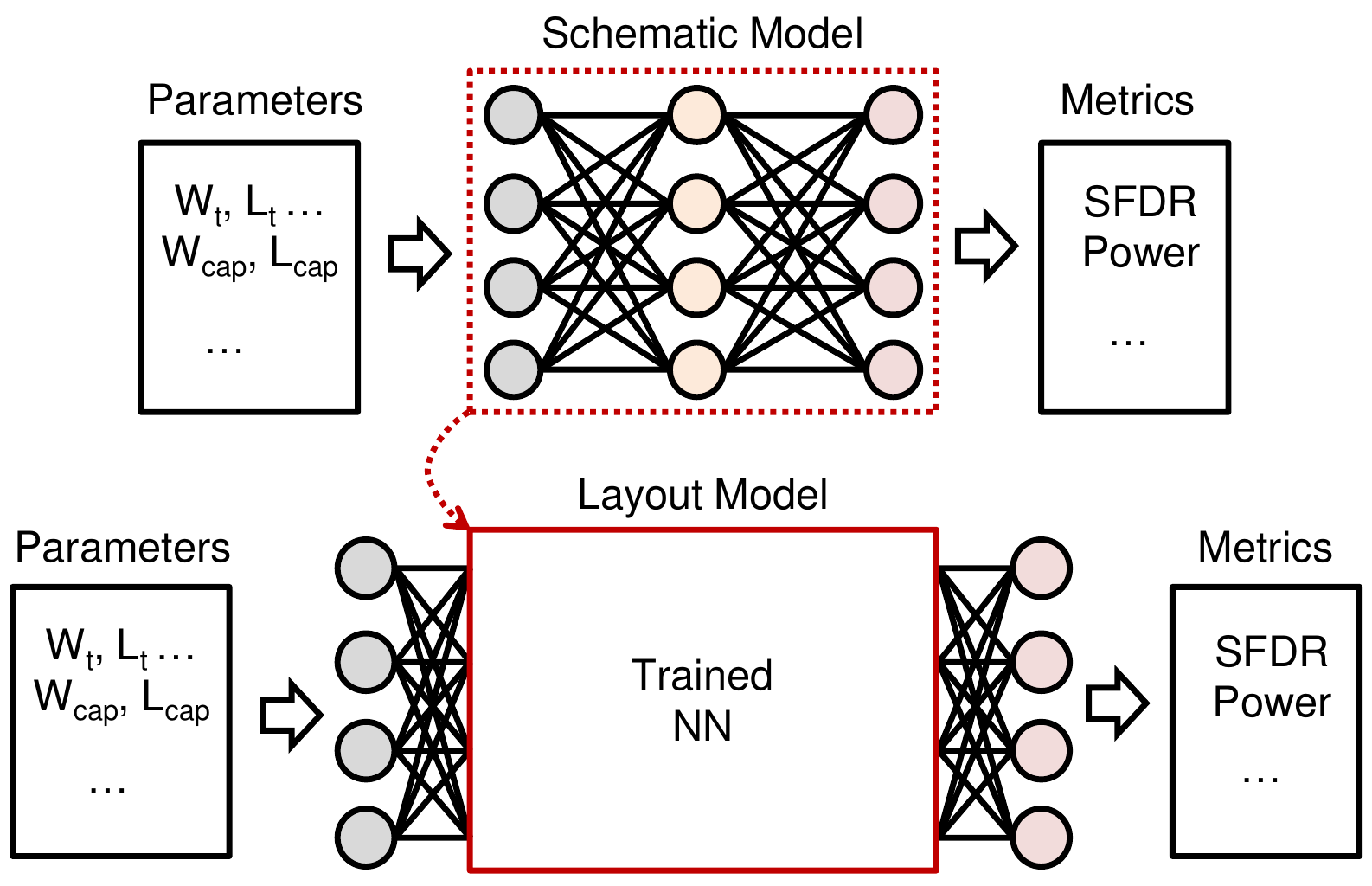}
    \caption{Transfer learning from schematic to layout models.}
    \label{TL}
\end{figure}

\subsection{Transfer Learning}

Despite the promising efficiency and accuracy of the approaches, most works mentioned above only focus on the schematic design in a particular technology node without considering PVT variations. To leverage the trained surrogate model when the design conditions are changed, \cite{liu2020tl} proposed a transfer learning (TL) technique. Instead of training the NN model from scratch with a large number of samples from the time-consuming post-layout simulations, the TL technique starts from an existing schematic-level circuit model, attaches one input linear layer and one output linear layer to the trained model, and only trains the new layers with a few post-layout samples. For the first time, \cite{liu2020tl} efficiently incorporated the layout parasitic information into the circuit surrogate model. Proved by experiments, this modeling method can effectively reduce the required training samples for a layout-level circuit model while maintaining a high modeling accuracy.

With this highly-efficient approach, \cite{su2022tafa} has successfully demonstrated a layout-aware AMS design flow from specification to layout, using an AMS filter as the test vehicle. \cite{liu2021tl} took one step further and applied TL to train a silicon-level circuit model and design the circuit incorporating both layout- and silicon-level information. This way, the NN-based approach for sophisticated AMS design has been significantly enhanced. Details of those design examples will be discussed in the next section.

\begin{figure}[!t]
    \centering
    \includegraphics[width=0.35\textwidth]{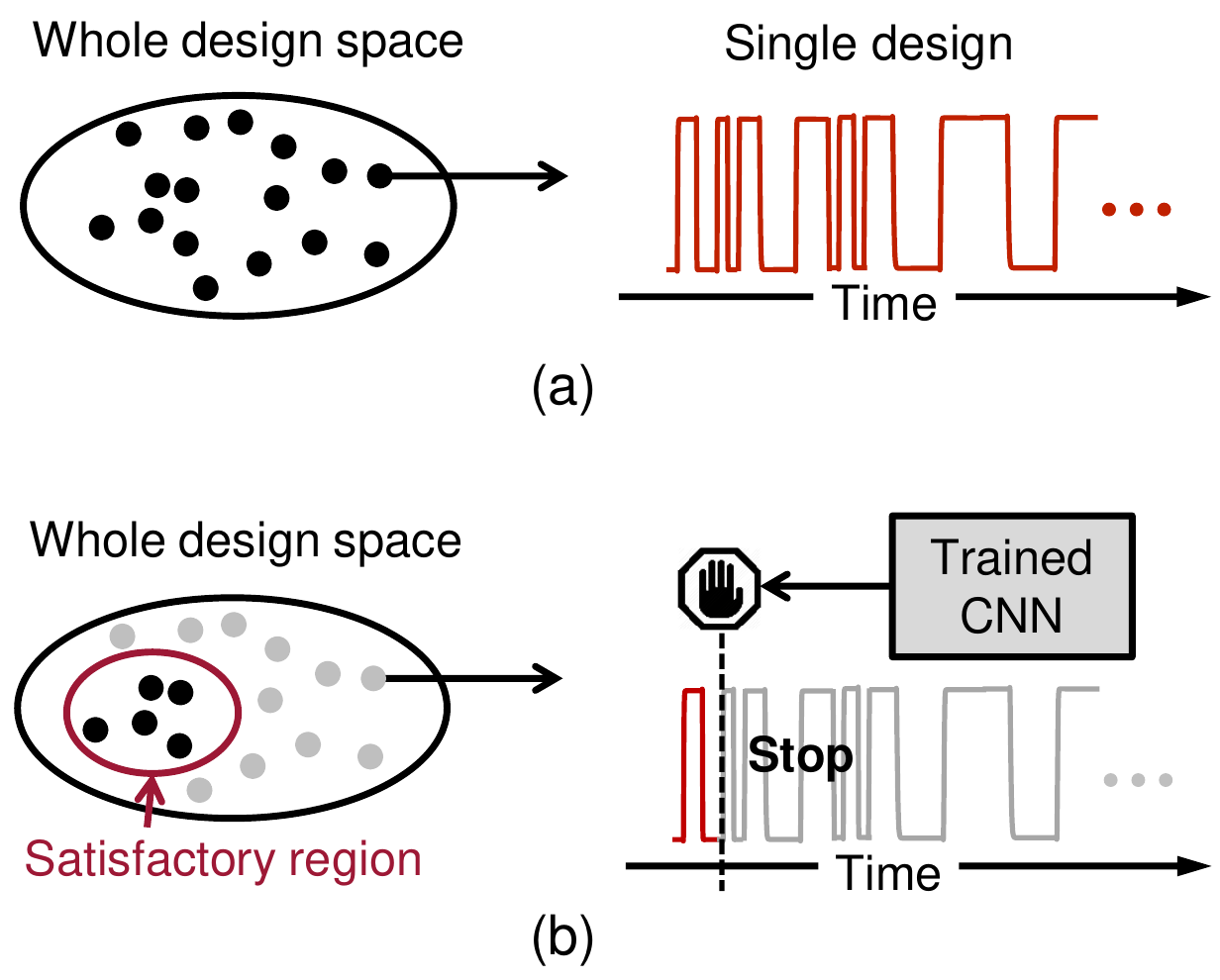}
    \caption{Training data generation with (a) typical transient simulation; (b) CEPA.}
    \label{CEPA}
\end{figure}

\subsection{Verification}
SPICE simulation plays an important role in the AMS circuit synthesis. For example, one would rely on accurate simulation results for validating the synthesized circuit. Unfortunately, AMS circuit simulations, especially transient simulations, are typically time-consuming because of the inherent complexity of the SPICE models and the required number of samples for FFT evaluation. To address these limitations, simulations of unsatisfactory designs can be terminated according to early-stage simulation results, which can potentially save a significant amount of machine computation time. Some physical and empirical formulas can quickly estimate the performance but lack high accuracy of judgment. \cite{zhang2020cepa} proposed a convolutional neural network (CNN) based early performance assertion scheme, named CEPA, for fast and accurate verification. CEPA takes a short duration of a transient waveform to predict the satisfaction of the target specifications, which are typically obtained in the frequency domain after long transient simulations. Trained with a few samples, the CNN can extract both human-recognizable and -unrecognizable features from the short transient waveform and use such features for performance prediction. Note that the learned features from the schematic simulations can be transferred to the post-layout model, with only a small number of training data from the post-layout simulation. As an application, CEPA can quickly narrow down the feasible design parameter space, which helps to sample the training data for the NN-based surrogate model and hence expedite the whole parameter search process.


\section{Proposed Design Flow and Examples for AMS Synthesis}

\subsection{Analog/Mixed-signal Parameter Search Engine}

\begin{figure}[!t]
    \centering
    \includegraphics[width=0.49\textwidth]{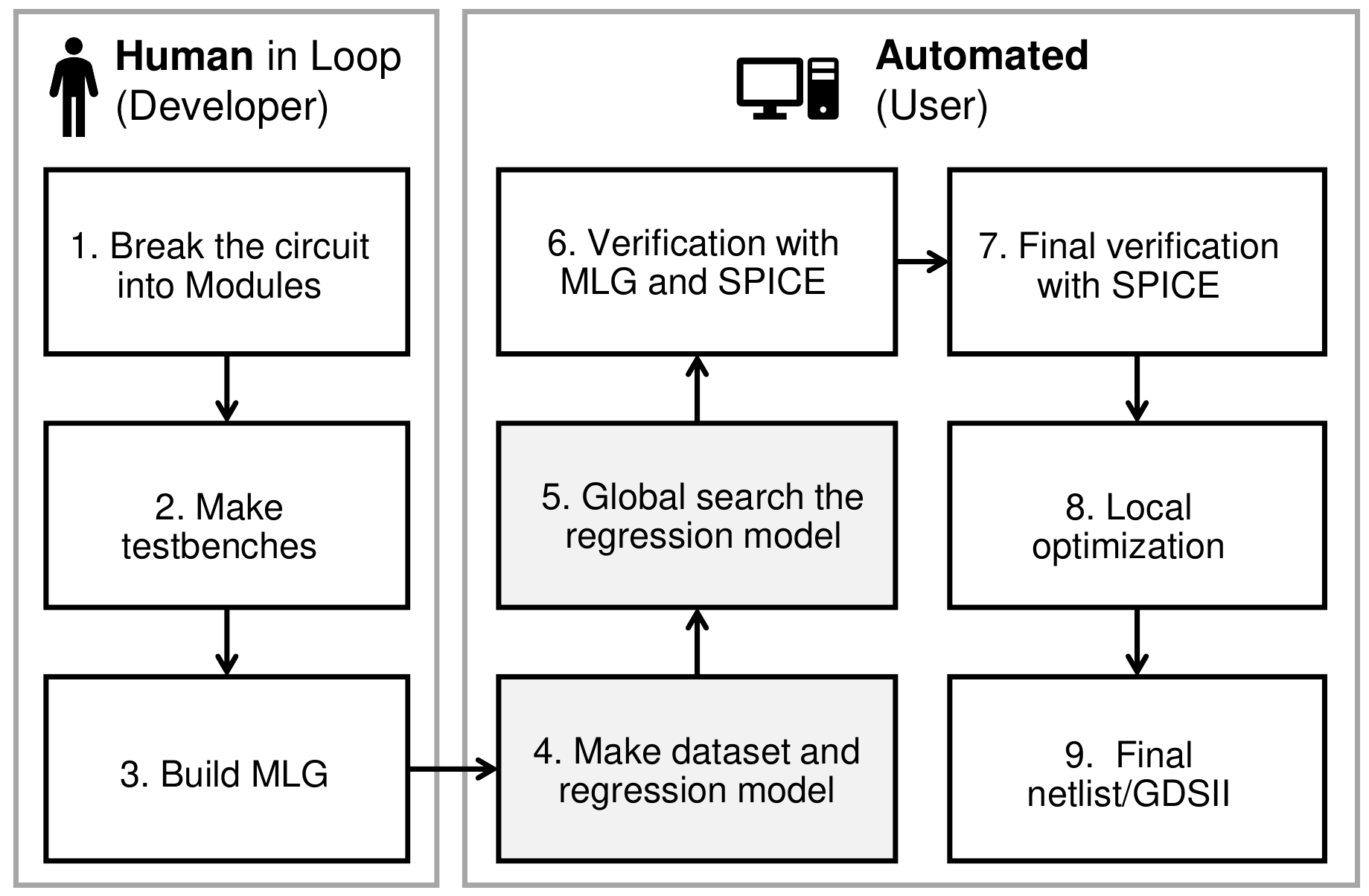}
    \caption{Proposed design flow based on AMPSE.}
    \label{AMPSE}
\end{figure}

Fig.~\ref{AMPSE} shows the proposed design flow based on an open-source AMS circuit generator, called Analog/Mixed-Signal Parameter Search Engine (AMPSE) \cite{AMPSE_website,hassanpourghadi2019automated}. First, AMPSE developers select promising circuit architectures from known good designs (KGD), break them into smaller modules, and parameterize the modules. Then, the developers make testbenches for characterizing each module and build MLG based on the connection between the modules. After the preparation, modeling and parameter search can be fully automated without human in the loop. NN serves as the surrogate model to represent the mapping between the design parameters and performance metrics. The model is trained with a dataset generated from the SPICE simulation, which is assisted by CEPA for reduced training efforts. Transfer learning is applied to incorporate post-layout information for improving the modeling accuracy. When the surrogate models of all the modules are prepared, they are used for global parameter search by connecting the models using MLG and applying gradient-based search algorithms. Thanks to the fast inference of NN, the search process is accelerated by orders of magnitude compared to the SPICE simulation based global search. AMPSE also suggests local optimization with SPICE model to fine-tune the circuit performance. Owing to the decent accuracy achieved by the surrogate model, the optimum design can be expected near the parameter candidates from the global search stage. Hence, the local optimization requires only a small number of iterations. For final verification, the SPICE simulation with the combined netlist is performed in the end. The whole AMPSE flow leverages both designer's knowledge and the recent advancement in machine learning and optimization, demonstrating highly automated and fast AMS circuit generation with a wide specification range and high performance.

\subsection{Example 1: SAR ADC}

\begin{figure}[!t]
    \centering
    \includegraphics[width=0.42\textwidth]{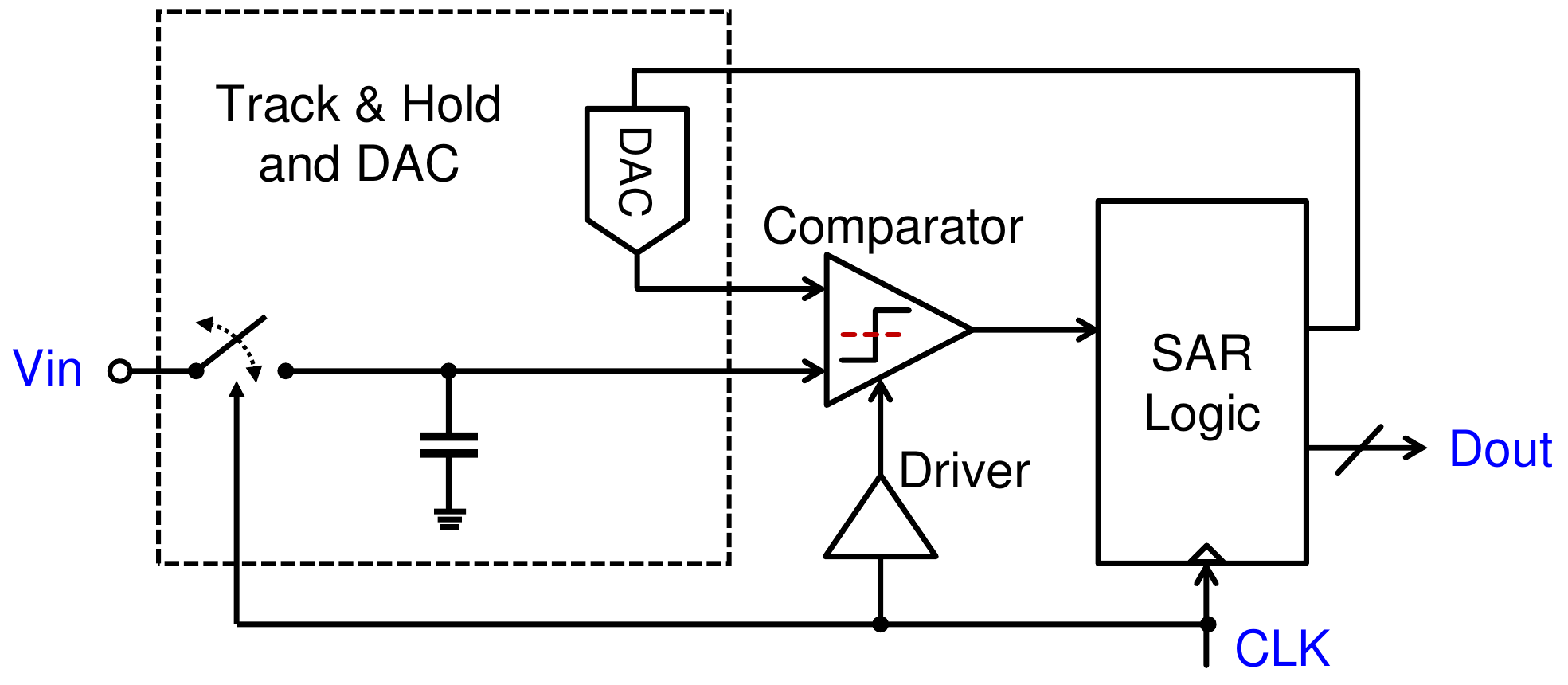}
    \caption{SAR ADC schematic.}
    \label{sar_sch}
\end{figure}

\begin{figure}[!t]
    \centering
    \includegraphics[width=0.48\textwidth]{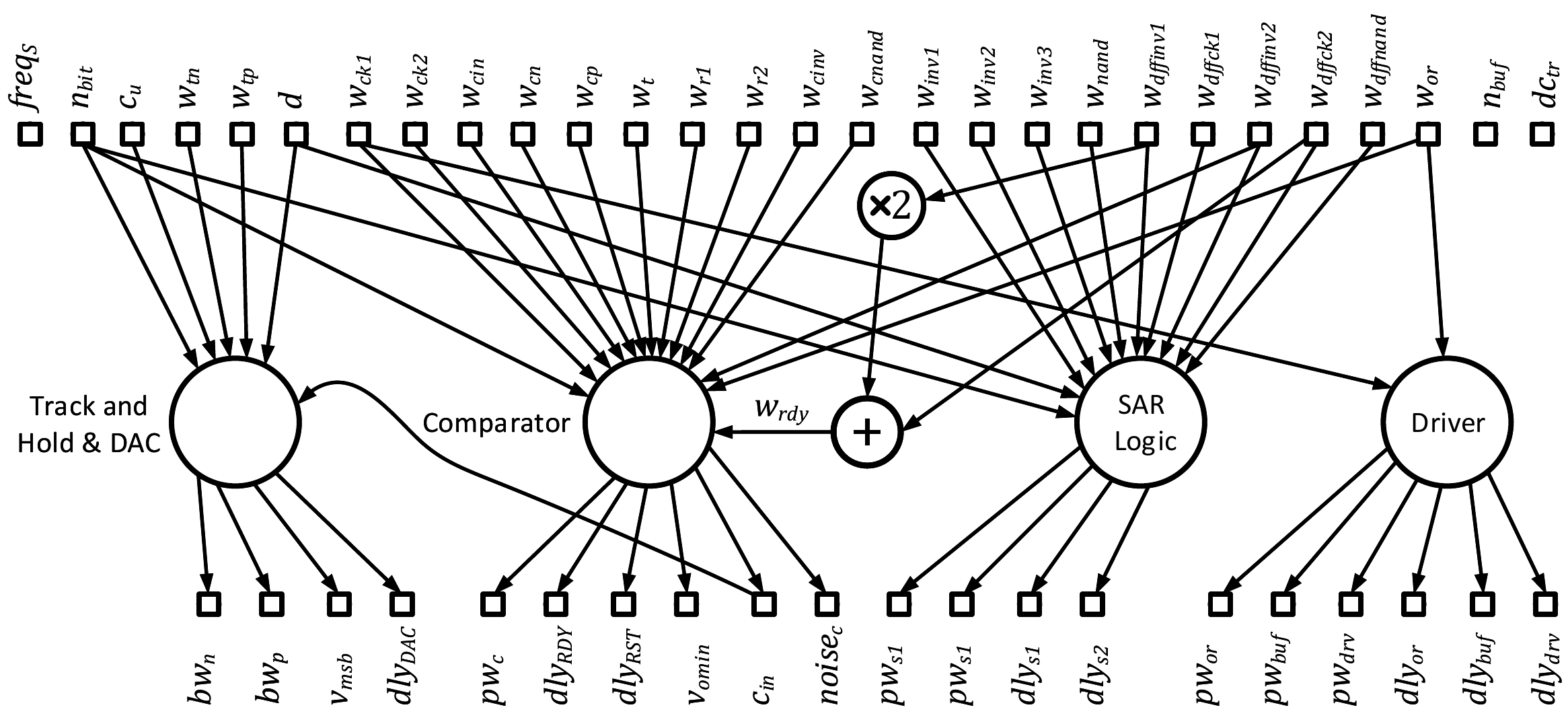}
    \caption{MLG of the SAR ADC.}
    \label{sar_graph}
\end{figure}

\begin{figure}[!t]
    \centering
    \includegraphics[width=0.46\textwidth]{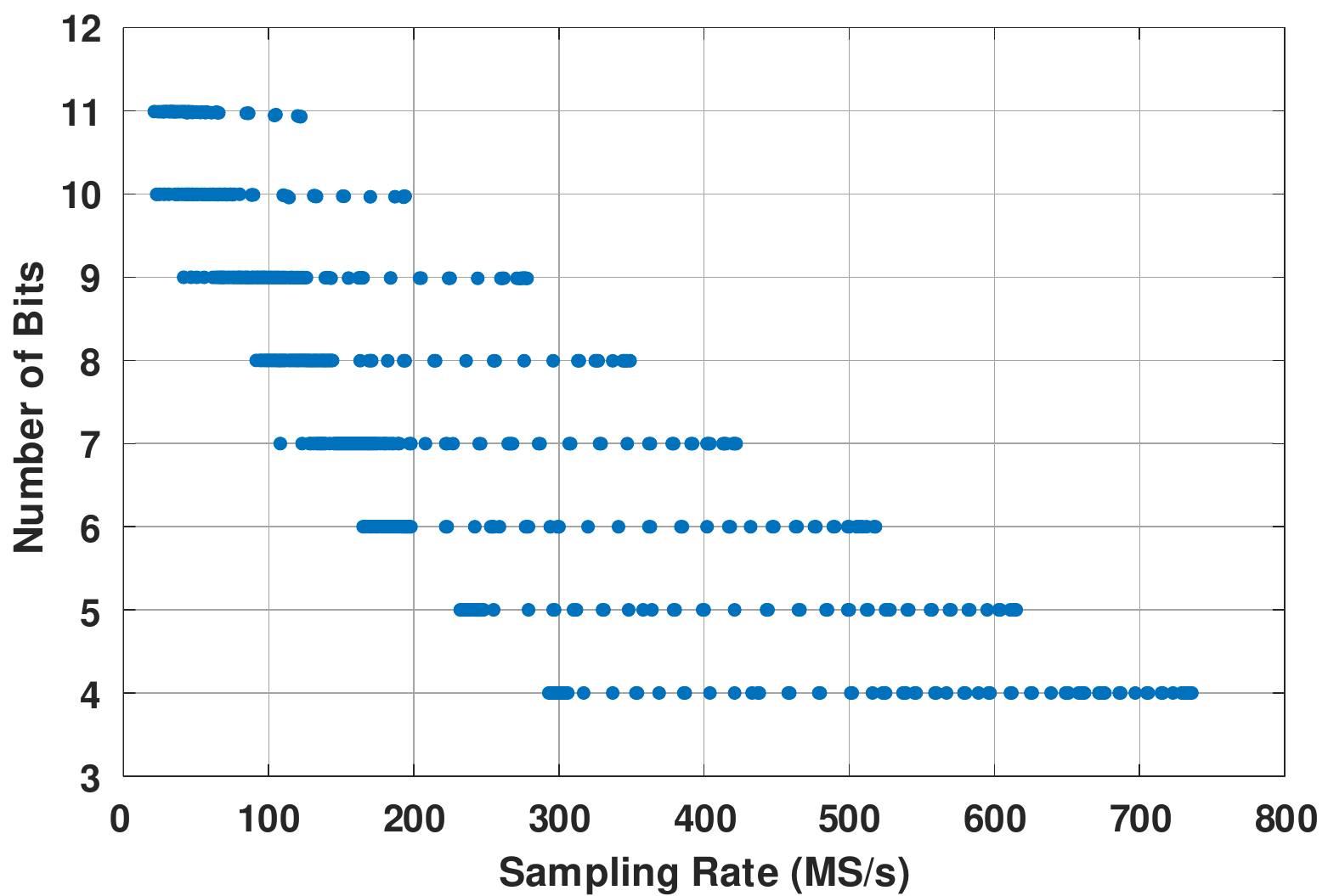}
    \caption{Number of bits versus sampling rate.}
    \label{banana_curve}
\end{figure}

In this example, the design was a SAR ADC from \cite{AMPSE_core_2021}. As shown in Fig.~\ref{sar_sch}, the ADC consists of four modules, i.e., track/hold and DAC, comparator, SAR logic, and a driver. There were 26 design parameters and 5 design specs. The objective was to satisfy all the specs while minimizing the power consumption. Fig.~\ref{sar_graph} shows the MLG of the SAR ADC, where the shared edges among modules represent the interface elements. AMPSE could generate around 500 different design candidates within 7 minutes which satisfied the specs. Fig.~\ref{banana_curve} shows the corresponding "banana" curve of the SAR ADC obtained by AMPSE. The plot depicts the possibility of the design outcomes for a given numbers of bits and sample rates. For a 6-bit 500 MS/s case, AMPSE reached similar performance as global search using the SPICE model while achieving almost 700 times faster search speed than the simulation-based method.

\subsection{Example 2: Delta-Sigma and RF DACs}

In \cite{zhang2020cepa}, we demonstrated the design of delta-sigma DAC in 65nm CMOS technology. The capacitor delta-sigma DAC consists of one inverter-based driver, one capacitor, and associated digital circuits. We first utilized CEPA to rapidly explore the design parameter space of the DAC and locate the feasible region as the target design space. We then used NN to model the mapping between the design parameters and the performance metrics within this design space and applied TL with post-layout simulation results to improve the model. Finally, we applied gradient descent on the NN model to search for the best possible design parameter combinations given the specifications. The DAC layout was generated using a mixed-signal layout flow \cite{zhang2020cepa}. The fully synthesized delta-sigma DAC achieved a 81 dB SFDR and 8.8-bit ENOB for 10 MHz signal bandwidth.

We have also explored RF-DAC-based AMS filter using time-approximation filter (TAF) architecture \cite{su2020taf}. The filter mainly consists of eight-channel time-interleaved RF DACs and a TAF pattern control circuits. We synthesized the control circuits using standard digital design flow and the DACs using a custom mixed-signal layout flow. The custom flow incorporated the designer's insights, such as symmetry and dummy constraints, to ensure high performance. A top-level script then integrated the two parts. To derive a nearly optimum filter response for the TAF, the impulse response was first designed based on the mathematical analysis and then optimized with a coordinate descent algorithm. This hybrid approximation scheme significantly reduced the time approximation errors of TAF over a wide range of filter's specifications.

\subsection{Example 3: Silicon verified and enhanced VCO design}

In the last design example, we demonstrated a ``from specification to silicon" design of voltage-controlled oscillators \cite{liu2021tl}. After training the schematic-level VCO model, we generated the layout samples using the ALIGN layout automation tool \cite{kunal2019align} and developed the layout-level model via TL. With the layout-level model, we designed ten different VCOs via AMPSE, laid out and taped out the silicon chip in the 12nm FinFET technology. The fabricated VCOs were measured and evaluated in terms of oscillation frequency and power consumption at different control voltages. Compared to the layout-level design results, the silicon measurement results showed a 12\% mean square variation. We then performed TL to tune the model using silicon-level samples (i.e., measurement data) and used the updated model to re-design the VCOs. Thanks to the silicon-level circuit model, the design flow could accurately predict the real silicon performance and found the corresponding design parameters with a 3.9\% mean square prediction error.


\begin{figure*}[!t]
    \centering
    \includegraphics[width=0.8\textwidth]{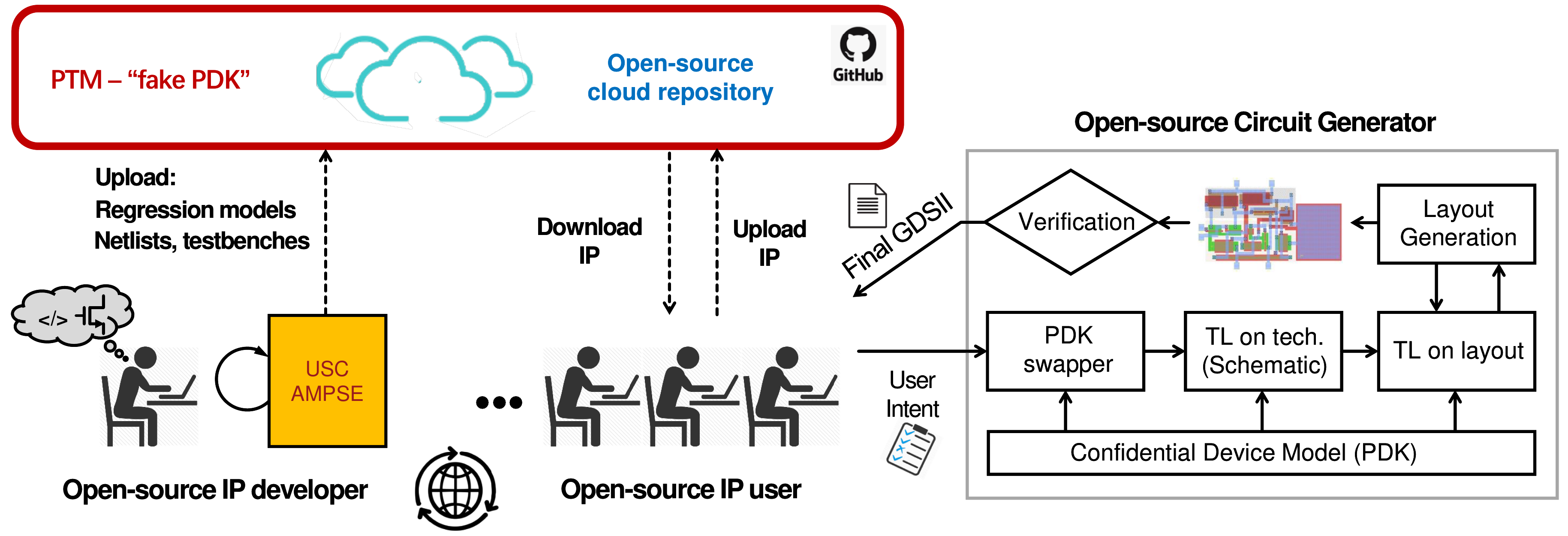}
    \caption{Open-source AMS design ecosystem.}
    \label{eco-system}
\end{figure*}

\section{Open-Source Ecosystem for AMS Design}

Moving forward, the growing demands and design cost of AMS circuits continuously challenge circuit designers and EDA tool developers. Besides, it is well-known that AMS circuit design is a highly specialized research area, where experienced designers' knowledge and intuition play a key role in successful designs. The shortage of design expertise becomes the bottleneck of the current design capacity of the industry. The recent DARPA Posh Open Source Hardware (POSH) program aims for an open-source hardware IP ecosystem. The open-source environment is widely adopted in software and digital design. However, it is still a fresh concept in the AMS circuit community for its technology dependency, IP sensitivity, reliability requirement etc. In Fig. \ref{eco-system}, we present a potential AMS circuit design ecosystem for sustainable and secure IP sharing, aiming to dramatically increase the AMS design capacity. The open-source AMS IP developers choose the silicon-proven circuit architectures and conduct the AMPSE flow to generate the surrogate models for a relatively large design parameter space. To avoid leaking out the confidential device model information, the developers should use the open-source predictive technology model (PTM) \cite{PTM_website} instead of the commercial process design kit (PDK) model for all the shared designs. The parameterized netlist, associated testbenches, and the surrogate models are all shared on the cloud. The IP developers can also upload verified design netlists with fixed parameters as KGDs. The IP users follow the procedure to obtain the desired AMS design:
\begin{enumerate}
    \item Download the target IP netlist from the cloud.
    \item Replace the PTM model used in the netlist with the actual PDK model.
    \item Apply TL to obtain an accurate surrogate circuit model.
    \item Use the surrogate model to find the circuit parameters that satisfy their design specifications. 
\end{enumerate}
IP users can also be developers by uploading the silicon/post-layout verified designs to the cloud repository. The kick-off of the open-source ecosystem can potentially lower the cost of AMS circuit development and promote more research outcomes to commercial products. Moreover, it can enable more complex AMS system innovation and integration that a single organization can never achieve.

\section{Conclusion}

As the technology scaling no longer leads to cost reduction and significant circuit performance improvement, AMS design automation has been gaining increasing attention from both industry and academia. This paper discussed the two main thrusts of current AMS circuit synthesis: (1) AMS circuit evolution towards mostly digital architecture and (2) ongoing application of machine learning algorithms in EDA tools. After reviewing the pros and cons of the state-of-the-art approaches, a complete AMS design flow based on NN surrogate model has been presented with examples.

 
 \section*{Acknowledgment}
The work is supported in part by DARPA ERI POSH program under Grant FA8650-18-2-7853 and in part by GlobalFoundries. The authors would like to thank Prof. Anthony F. J. Levi and Prof. Sandeep K. Gupta from the University of Southern California for technical discussions. 
 
\input{mybib.bbl}

\bibliographystyle{IEEEtran}
\bibliography{mybib}

\end{document}

%% file: mybib.bbl

%% file: main.bbl
\begin{thebibliography}{10}
\providecommand{\url}[1]{#1}
\csname url@samestyle\endcsname
\providecommand{\newblock}{\relax}
\providecommand{\bibinfo}[2]{#2}
\providecommand{\BIBentrySTDinterwordspacing}{\spaceskip=0pt\relax}
\providecommand{\BIBentryALTinterwordstretchfactor}{4}
\providecommand{\BIBentryALTinterwordspacing}{\spaceskip=\fontdimen2\font plus
\BIBentryALTinterwordstretchfactor\fontdimen3\font minus
  \fontdimen4\font\relax}
\providecommand{\BIBforeignlanguage}[2]{{%
\expandafter\ifx\csname l@#1\endcsname\relax
\typeout{** WARNING: IEEEtran.bst: No hyphenation pattern has been}%
\typeout{** loaded for the language `#1'. Using the pattern for}%
\typeout{** the default language instead.}%
\else
\language=\csname l@#1\endcsname
\fi
#2}}
\providecommand{\BIBdecl}{\relax}
\BIBdecl

\bibitem{kahng2018new}
A.~B. Kahng, ``New directions for learning-based {IC} design tools and
  methodologies,'' in \emph{2018 23rd Asia and South Pacific Design Automation
  Conference (ASP-DAC)}.\hskip 1em plus 0.5em minus 0.4em\relax IEEE, 2018, pp.
  405--410.

\bibitem{manganaro2018emerging}
G.~Manganaro, ``Emerging data converter architectures and techniques,'' in
  \emph{2018 IEEE Custom Integrated Circuits Conference (CICC)}.\hskip 1em plus
  0.5em minus 0.4em\relax IEEE, 2018, pp. 1--8.

\bibitem{robertson2016data}
D.~Robertson, A.~Buchwald, M.~Flynn, H.-S. Lee, U.-K. Moon, and B.~Murmann,
  ``Data converter reflections: 19 papers from the last ten years that deserve
  a second look,'' in \emph{ESSCIRC Conference 2016: 42nd European Solid-State
  Circuits Conference}.\hskip 1em plus 0.5em minus 0.4em\relax IEEE, 2016, pp.
  161--164.

\bibitem{loke2018analog}
A.~L. Loke, D.~Yang, T.~T. Wee, J.~L. Holland, P.~Isakanian, K.~Rim, S.~Yang,
  J.~S. Schneider, G.~Nallapati, S.~Dundigal \emph{et~al.},
  ``Analog/mixed-signal design challenges in 7-nm {CMOS} and beyond,'' in
  \emph{2018 IEEE Custom Integrated Circuits Conference (CICC)}.\hskip 1em plus
  0.5em minus 0.4em\relax IEEE, 2018, pp. 1--8.

\bibitem{mitola1995software}
J.~Mitola, ``The software radio architecture,'' \emph{IEEE Communications
  magazine}, vol.~33, no.~5, pp. 26--38, 1995.

\bibitem{murmann200312}
B.~Murmann and B.~E. Boser, ``A 12-bit 75-{MS}/s pipelined {ADC} using
  open-loop residue amplification,'' \emph{IEEE Journal of Solid-State
  Circuits}, vol.~38, no.~12, pp. 2040--2050, 2003.

\bibitem{chiu2004least}
Y.~Chiu, C.~W. Tsang, B.~Nikolic, and P.~R. Gray, ``Least mean square adaptive
  digital background calibration of pipelined analog-to-digital converters,''
  \emph{IEEE Transactions on Circuits and Systems I: Regular Papers}, vol.~51,
  no.~1, pp. 38--46, 2004.

\bibitem{ali202012}
A.~M. Ali, H.~Dinc, P.~Bhoraskar, S.~Bardsley, C.~Dillon, M.~McShea, J.~P.
  Periathambi, and S.~Puckett, ``A 12-b 18-{GS}/s {RF} sampling {ADC} with an
  integrated wideband track-and-hold amplifier and background calibration,''
  \emph{IEEE Journal of Solid-State Circuits}, vol.~55, no.~12, pp. 3210--3224,
  2020.

\bibitem{su201612}
S.~{Su} and M.~S.-W. {Chen}, ``A 12-bit 2 {GS}/s dual-rate hybrid {DAC} with
  pulse-error pre-distortion and in-band noise cancellation achieving $>$ 74
  d{B}c {SFDR} and $<$-80 d{B}c {IM}3 up to 1 {GH}z in 65 nm {CMOS},''
  \emph{IEEE Journal of Solid-State Circuits}, vol.~51, no.~12, pp. 2963--2978,
  2016.

\bibitem{lin201816b}
C.-H. Lin, K.~L.~J. Wong, T.-Y. Kim, G.~R. Xie, D.~Major, G.~Unruh, S.~R.
  Dommaraju, H.~Eberhart, and A.~Venes, ``A 16b 6{GS}/s nyquist {DAC} with
  {IMD} $<$-90d{B}c up to 1.9 {GH}z in 16nm {CMOS},'' in \emph{2018 IEEE
  International Solid-State Circuits Conference-(ISSCC)}.\hskip 1em plus 0.5em
  minus 0.4em\relax IEEE, 2018, pp. 360--362.

\bibitem{su201816}
S.~Su and M.~S.-W. Chen, ``A 16-bit 12-{GS}/s single-/dual-rate {DAC} with a
  successive bandpass delta-sigma modulator achieving $<$-67-d{B}c {IM}3 within
  {DC} to 6-{GH}z tunable passbands,'' \emph{IEEE Journal of Solid-State
  Circuits}, vol.~53, no.~12, pp. 3517--3527, 2018.

\bibitem{wu20162}
J.~Wu, G.~Cusmai, A.~Wei-Te~Chou, T.~Wang, B.~Shen, V.~Periasamy, M.-H. Hsieh,
  C.-Y. Chen, L.~He, L.~K. Tan \emph{et~al.}, ``A 2.7 m{W}/channel 48--1000
  {MH}z direct sampling full-band cable receiver,'' \emph{IEEE Journal of
  Solid-State Circuits}, vol.~51, no.~4, pp. 845--859, 2016.

\bibitem{spiridon2013375}
S.~Spiridon, J.~van~der Tang, H.~Yan, H.-F. Chen, D.~Guermandi, X.~Liu,
  E.~Arslan, F.~van~der Goes, and K.~Bult, ``A 375 m{W} multimode {DAC}-based
  transmitter with 2.2 {GH}z signal bandwidth and in-band {IM}3 $<$58 d{B}c in
  40 nm {CMOS},'' \emph{IEEE journal of solid-state circuits}, vol.~48, no.~7,
  pp. 1595--1604, 2013.

\bibitem{staszewski2005all}
R.~B. Staszewski, J.~L. Wallberg, S.~Rezeq, C.-M. Hung, O.~E. Eliezer, S.~K.
  Vemulapalli, C.~Fernando, K.~Maggio, R.~Staszewski, N.~Barton \emph{et~al.},
  ``All-digital {PLL} and transmitter for mobile phones,'' \emph{IEEE Journal
  of Solid-State Circuits}, vol.~40, no.~12, pp. 2469--2482, 2005.

\bibitem{deng2014fully}
W.~Deng, D.~Yang, T.~Ueno, T.~Siriburanon, S.~Kondo, K.~Okada, and
  A.~Matsuzawa, ``A fully synthesizable all-digital {PLL} with interpolative
  phase coupled oscillator, current-output {DAC}, and fine-resolution digital
  varactor using gated edge injection technique,'' \emph{IEEE Journal of
  Solid-State Circuits}, vol.~50, no.~1, pp. 68--80, 2014.

\bibitem{ho2016fractional}
C.-R. Ho and M.~S.-W. Chen, ``A fractional-{N} {DPLL} with calibration-free
  multi-phase injection-locked {TDC} and adaptive single-tone spur cancellation
  scheme,'' \emph{IEEE Transactions on Circuits and Systems I: Regular Papers},
  vol.~63, no.~8, pp. 1111--1122, 2016.

\bibitem{chen2010calibration}
M.~S.-W. Chen, D.~Su, and S.~Mehta, ``A calibration-free 800 {MH}z
  fractional-{N} digital {PLL} with embedded {TDC},'' \emph{IEEE Journal of
  Solid-State Circuits}, vol.~45, no.~12, pp. 2819--2827, 2010.

\bibitem{ho2016digital}
C.-R. Ho and M.~S.-W. Chen, ``A digital {PLL} with feedforward multi-tone spur
  cancellation scheme achieving $<$--73 d{B}c fractional spur and $<$--110
  d{B}c reference spur in 65 nm {CMOS},'' \emph{IEEE Journal of Solid-State
  Circuits}, vol.~51, no.~12, pp. 3216--3230, 2016.

\bibitem{zhang2021fractional}
Q.~Zhang, S.~Su, C.-R. Ho, and M.~S.-W. Chen, ``A fractional-{N} digital {MDLL}
  with background two-point {DTC} calibration,'' \emph{IEEE Journal of
  Solid-State Circuits}, 2021.

\bibitem{kundu2020self}
S.~Kundu, L.~Chai, K.~Chandrashekar, S.~Pellerano, and B.~R. Carlton, ``A
  self-calibrated 2-bit time-period comparator-based synthesized fractional-{N}
  {MDLL} in 22-nm {F}in{FET} {CMOS},'' \emph{IEEE Journal of Solid-State
  Circuits}, vol.~56, no.~1, pp. 43--54, 2020.

\bibitem{okuma2010cicc}
Y.~Okuma, K.~Ishida, Y.~Ryu, X.~Zhang, P.-H. Chen, K.~Watanabe, M.~Takamiya,
  and T.~Sakurai, ``0.5-{V} input digital {LDO} with 98.7\% current efficiency
  and 2.7-$\mu${A} quiescent current in 65nm {CMOS},'' in \emph{IEEE Custom
  Integrated Circuits Conference 2010}.\hskip 1em plus 0.5em minus 0.4em\relax
  IEEE, 2010, pp. 1--4.

\bibitem{su201512}
S.~Su, T.-I. Tsai, P.~K. Sharma, and M.~S.-W. Chen, ``A 12 bit 1 {GS}/s
  dual-rate hybrid {DAC} with an 8 {GS}/s unrolled pipeline delta-sigma
  modulator achieving $>$ 75 d{B} {SFDR} over the nyquist band,'' \emph{IEEE
  Journal of Solid-State Circuits}, vol.~50, no.~4, pp. 896--907, 2015.

\bibitem{chen20066}
S.-W.~M. Chen and R.~W. Brodersen, ``A 6-bit 600-{MS}/s 5.3-m{W} asynchronous
  {ADC} in 0.13-$\mu$m {CMOS},'' \emph{IEEE Journal of Solid-State Circuits},
  vol.~41, no.~12, pp. 2669--2680, 2006.

\bibitem{ding2018circuit}
M.~Ding, G.~Chen, P.~Harpe, B.~Busze, Y.-H. Liu, C.~Bachmann, K.~Philips, and
  A.~van Roermund, ``A circuit-design-driven tool with a hybrid automation
  approach for {SAR ADC}s in {I}o{T},'' in \emph{2018 Design, Automation \&
  Test in Europe Conference \& Exhibition (DATE)}.\hskip 1em plus 0.5em minus
  0.4em\relax IEEE, 2018, pp. 672--675.

\bibitem{chang2018bag2}
E.~Chang, J.~Han, W.~Bae, Z.~Wang, N.~Narevsky, B.~Nikolic, and E.~Alon,
  ``{BAG}2: A process-portable framework for generator-based {AMS} circuit
  design,'' in \emph{2018 IEEE Custom Integrated Circuits Conference
  (CICC)}.\hskip 1em plus 0.5em minus 0.4em\relax IEEE, 2018, pp. 1--8.

\bibitem{dudek2000high}
P.~Dudek, S.~Szczepanski, and J.~V. Hatfield, ``A high-resolution cmos
  time-to-digital converter utilizing a vernier delay line,'' \emph{IEEE
  Journal of Solid-State Circuits}, vol.~35, no.~2, pp. 240--247, 2000.

\bibitem{mohsen2018vcoadc}
M.~Hassanpourghadi, P.~K. Sharma, and M.~S.-W. Chen, ``A 6-b, 800-{MS}/s,
  3.62-m{W} nyquist rate {AC}-coupled {VCO}-based {ADC} in 65-nm {CMOS},''
  \emph{IEEE Transactions on Circuits and Systems I: Regular Papers}, vol.~64,
  no.~6, pp. 1354--1367, 2017.

\bibitem{jssc_vcobased5g_2019}
M.~Baert and W.~Dehaene, ``A 5-{GS}/s 7.2-{ENOB} {T}ime-{I}nterleaved
  {VCO}-{B}ased {ADC} {A}chieving 30.5 fj/cs,'' \emph{IEEE Journal of
  Solid-State Circuits}, vol.~55, no.~6, pp. 1577--1587, 2020.

\bibitem{jssc_rnstadc_2018}
S.~Zhu, B.~Wu, Y.~Cai, and Y.~Chiu, ``A 2-{GS}/s 8-bit {N}on-{I}nterleaved
  {T}ime-{D}omain {F}lash {ADC} {B}ased on {R}emainder {N}umber {S}ystem in
  65-nm {CMOS},'' \emph{IEEE Journal of Solid-State Circuits}, vol.~53, no.~4,
  pp. 1172--1183, 2018.

\bibitem{mohsen2019ppstdc}
M.~Hassanpourghadi and M.~S.-W. Chen, ``A 2-way 7.3-bit 10 {GS}/s time-based
  folding {ADC} with passive pulse-shrinking cells,'' in \emph{2019 IEEE Custom
  Integrated Circuits Conference (CICC)}, 2019, pp. 1--4.

\bibitem{Juzheng2022Timebase}
J.~Liu, M.~Hassanpourghadi, and M.~S.-W. Chen, ``A 10{GS}/s 8b 25fj/c-s
  2850$um^2$ {T}wo-{S}tep {T}ime-domain {ADC} {U}sing {D}elay-{T}racking
  {P}ipelined-{SAR} {TDC} with 500fs {T}ime {S}tep in 14nm {CMOS}
  {T}echnology,'' in \emph{2022 IEEE International Solid - State Circuits
  Conference - (ISSCC)}.

\bibitem{xu2017scaling}
B.~Xu, S.~Li, N.~Sun, and D.~Z. Pan, ``A scaling compatible, synthesis friendly
  {VCO}-based delta-sigma {ADC} design and synthesis methodology,'' in
  \emph{2017 54th ACM/EDAC/IEEE Design Automation Conference (DAC)}.\hskip 1em
  plus 0.5em minus 0.4em\relax IEEE, 2017, pp. 1--6.

\bibitem{chen2019digital}
Z.~Chen, H.~Zhou, and J.~Gu, ``Digital compatible synthesis, placement and
  implementation of mixed-signal time-domain computing,'' in \emph{2019 56th
  ACM/IEEE Design Automation Conference (DAC)}.\hskip 1em plus 0.5em minus
  0.4em\relax IEEE, 2019, pp. 1--6.

\bibitem{crovetti2013digital}
P.~S. Crovetti, ``A digital-based analog differential circuit,'' \emph{IEEE
  Transactions on Circuits and Systems I: Regular Papers}, vol.~60, no.~12, pp.
  3107--3116, 2013.

\bibitem{Liu2015cicc}
J.~{Liu}, A.~{Fahmy}, T.~{Kim}, and N.~{Maghari}, ``A fully synthesized 0.4{V}
  77d{B} {SFDR} reprogrammable {SRMC} filter using digital standard cells,'' in
  \emph{2015 IEEE Custom Integrated Circuits Conference (CICC)}.\hskip 1em plus
  0.5em minus 0.4em\relax IEEE, 2015, pp. 1--4.

\bibitem{su2020jssc}
S.~{Su} and M.~S.-W. {Chen}, ``A time-approximation filter for direct {RF}
  transmitter,'' \emph{IEEE Journal of Solid-State Circuits}, pp. 1--1, 2020.

\bibitem{su2020taf}
S.~Su and M.~S.-W. Chen, ``A {SAW}-less direct-digital {RF} modulator with
  tri-level time-approximation filter and reconfigurable dual-band delta-sigma
  modulation,'' in \emph{2020 IEEE International Solid-State Circuits
  Conference-(ISSCC)}.\hskip 1em plus 0.5em minus 0.4em\relax IEEE, 2020, pp.
  174--176.

\bibitem{wei2021analog}
P.-H. Wei and B.~Murmann, ``Analog and mixed-signal layout automation using
  digital place-and-route tools,'' \emph{IEEE Transactions on Very Large Scale
  Integration (VLSI) Systems}, vol.~29, no.~11, pp. 1838--1849, 2021.

\bibitem{su2022tafa}
S.~Su, Q.~Zhang, J.~Liu, M.~Hassanpourghadi, R.~Rasul, and M.~S.-W. Chen,
  ``{TAFA}: Design automation of analog mixed-signal {FIR} filters using time
  approximation architecture,'' in \emph{2022 27th Asia and South Pacific
  Design Automation Conference (ASP-DAC)}.\hskip 1em plus 0.5em minus
  0.4em\relax IEEE, 2022.

\bibitem{harjani1989oasys}
R.~Harjani, R.~A. Rutenbar, and L.~R. Carley, ``{OASYS}: A framework for analog
  circuit synthesis,'' \emph{IEEE Transactions on Computer-Aided Design of
  Integrated Circuits and Systems}, vol.~8, no.~12, pp. 1247--1266, 1989.

\bibitem{boyd2001optimal}
S.~P. Boyd, T.~H. Lee \emph{et~al.}, ``Optimal design of a {CMOS} op-amp via
  geometric programming,'' \emph{IEEE Transactions on Computer-aided design of
  integrated circuits and systems}, vol.~20, no.~1, pp. 1--21, 2001.

\bibitem{colleran2003optimization}
D.~M. Colleran, C.~Portmann, A.~Hassibi, C.~Crusius, S.~S. Mohan, S.~Boyd,
  T.~H. Lee, and M.~del Mar~Hershenson, ``Optimization of phase-locked loop
  circuits via geometric programming,'' in \emph{Proceedings of the IEEE 2003
  Custom Integrated Circuits Conference, 2003.}\hskip 1em plus 0.5em minus
  0.4em\relax IEEE, 2003, pp. 377--380.

\bibitem{del2002design}
M.~del Mar~Hershenson, ``Design of pipeline analog-to-digital converters via
  geometric programming,'' in \emph{Proceedings of the 2002 IEEE/ACM
  International Conference on Computer-Aided Design (ICCAD)}, 2002, pp.
  317--324.

\bibitem{de2005mixed}
F.~De~Bernardinis, P.~Nuzzo, and A.~S. Vincentelli, ``Mixed signal design space
  exploration through analog platforms,'' in \emph{Proceedings of the 42nd
  annual Design Automation Conference}, 2005, pp. 875--880.

\bibitem{kiely2004performance}
T.~Kiely and G.~Gielen, ``Performance modeling of analog integrated circuits
  using least-squares support vector machines,'' in \emph{Proceedings Design,
  Automation and Test in Europe Conference and Exhibition}, vol.~1.\hskip 1em
  plus 0.5em minus 0.4em\relax IEEE, 2004, pp. 448--453.

\bibitem{liu2002remembrance}
H.~Liu, A.~Singhee, R.~A. Rutenbar, and L.~R. Carley, ``Remembrance of circuits
  past: macromodeling by data mining in large analog design spaces,'' in
  \emph{Proceedings of the 39th annual Design Automation Conference}, 2002, pp.
  437--442.

\bibitem{wang2015bayesian}
F.~Wang, P.~Cachecho, W.~Zhang, S.~Sun, X.~Li, R.~Kanj, and C.~Gu, ``Bayesian
  model fusion: large-scale performance modeling of analog and mixed-signal
  circuits by reusing early-stage data,'' \emph{IEEE Transactions on
  Computer-Aided Design of Integrated Circuits and Systems}, vol.~35, no.~8,
  pp. 1255--1268, 2015.

\bibitem{lyu2017efficient}
W.~Lyu, P.~Xue, F.~Yang, C.~Yan, Z.~Hong, X.~Zeng, and D.~Zhou, ``An efficient
  bayesian optimization approach for automated optimization of analog
  circuits,'' \emph{IEEE Transactions on Circuits and Systems I: Regular
  Papers}, vol.~65, no.~6, pp. 1954--1967, 2017.

\bibitem{gielen2009tcad}
T.~McConaghy and G.~G.~E. Gielen, ``Template-free symbolic performance modeling
  of analog circuits via canonical-form functions and genetic programming,''
  \emph{IEEE Transactions on Computer-Aided Design of Integrated Circuits and
  Systems}, vol.~28, no.~8, pp. 1162--1175, 2009.

\bibitem{li2020tcad1}
Y.~{Li}, Y.~{Wang}, Y.~{Li}, R.~{Zhou}, and Z.~{Lin}, ``An artificial neural
  network assisted optimization system for analog design space exploration,''
  \emph{IEEE Transactions on Computer-Aided Design of Integrated Circuits and
  Systems}, vol.~39, no.~10, pp. 2640--2653, 2020.

\bibitem{wolfe2003tcad1}
G.~{Wolfe} and R.~{Vemuri}, ``Extraction and use of neural network models in
  automated synthesis of operational amplifiers,'' \emph{IEEE Transactions on
  Computer-Aided Design of Integrated Circuits and Systems}, vol.~22, no.~2,
  pp. 198--212, Feb 2003.

\bibitem{islam2019icsmas}
G.~{İslamoğlu}, T.~O. {Çakici}, E.~{Afacan}, and G.~{Dündar}, ``Artificial
  neural network assisted analog {IC} sizing tool,'' in \emph{2019 16th
  International Conference on Synthesis, Modeling, Analysis and Simulation
  Methods and Applications to Circuit Design (SMACD)}.\hskip 1em plus 0.5em
  minus 0.4em\relax Lausanne, Switzerland, Switzerland: IEEE, 2019, pp. 9--12.

\bibitem{garit2012icvd}
O.~{Garitselov}, S.~P. {Mohanty}, and E.~{Kougianos}, ``Fast-accurate
  non-polynomial metamodeling for nano-{CMOS} {PLL} design optimization,'' in
  \emph{2012 25th International Conference on VLSI Design}.\hskip 1em plus
  0.5em minus 0.4em\relax Hyderabad, India: IEEE, 2012, pp. 316--321.

\bibitem{modelbased_multiObj_2005}
T.~{Eeckelaert}, T.~{McConaghy}, and G.~{Gielen}, ``Efficient multiobjective
  synthesis of analog circuits using hierarchical {P}areto-optimal performance
  hypersurfaces,'' in \emph{Design, Automation and Test in Europe}.\hskip 1em
  plus 0.5em minus 0.4em\relax Munich, Germany, Germany: IEEE, March 2005, pp.
  1070--1075 Vol. 2.

\bibitem{lorenco2019icsmas}
N.~{Lourenço}, E.~{Afacan}, R.~{Martins}, F.~{Passos}, A.~{Canelas},
  R.~{Póvoa}, N.~{Horta}, and G.~{Dundar}, ``Using polynomial regression and
  artificial neural networks for reusable analog ic sizing,'' in \emph{2019
  16th International Conference on Synthesis, Modeling, Analysis and Simulation
  Methods and Applications to Circuit Design (SMACD)}.\hskip 1em plus 0.5em
  minus 0.4em\relax Lausanne, Switzerland: IEEE, 2019, pp. 13--16.

\bibitem{AMPSE_core_2021}
\BIBentryALTinterwordspacing
M.~Hassanpourghadi, R.~A. Rasul, and M.~S.-W. Chen, ``A module-linking graph
  assisted hybrid optimization framework for custom analog and mixed-signal
  circuit parameter synthesis,'' \emph{ACM Trans. Des. Autom. Electron. Syst.},
  vol.~26, no.~5, Jun. 2021. [Online]. Available:
  \url{https://doi.org/10.1145/3456722}
\BIBentrySTDinterwordspacing

\bibitem{mohsen2021acmdac}
M.~Hassanpourghadi, S.~Su, R.~A. Rasul, J.~Liu, Q.~Zhang, and M.~S.-W. Chen,
  ``Circuit connectivity inspired neural network for analog mixed-signal
  functional modeling,'' in \emph{Proceedings of the 58th ACM/EDAC/IEEE Design
  Automation Conference (DAC)}.\hskip 1em plus 0.5em minus 0.4em\relax IEEE,
  2021.

\bibitem{liu2020tl}
J.~Liu, M.~Hassanpourghadi, Q.~Zhang, S.~Su, and M.~S.-W. Chen, ``{T}ransfer
  learning with bayesian optimization-aided sampling for efficient {AMS}
  circuit modeling,'' in \emph{2020 IEEE/ACM International Conference on
  Computer-Aided Design (ICCAD)}.\hskip 1em plus 0.5em minus 0.4em\relax ACM,
  2020.

\bibitem{liu2021tl}
J.~Liu, S.~Su, M.~Madhusudan, M.~Hassanpourghadi, S.~Saunders, Q.~Zhang,
  R.~Rasul, Y.~Li, J.~Hu, A.~K. Sharma, S.~S. Sapatnekar, R.~Harjani, A.~Levi,
  S.~Gupta, and M.~S.-W. Chen, ``{F}rom specification to silicon: Towards
  analog/mixed-signal design automation using surrogate {NN} models with
  transfer learning,'' in \emph{2021 IEEE/ACM International Conference on
  Computer-Aided Design (ICCAD)}.\hskip 1em plus 0.5em minus 0.4em\relax ACM,
  2021.

\bibitem{zhang2020cepa}
Q.~Zhang, S.~Su, J.~Liu, and M.~S.-W. Chen, ``{CEPA}: {CNN}-based early
  performance assertion scheme for analog and mixed-signal circuit
  simulation,'' in \emph{2020 IEEE/ACM International Conference on
  Computer-Aided Design (ICCAD)}.\hskip 1em plus 0.5em minus 0.4em\relax ACM,
  2020.

\bibitem{AMPSE_website}
\BIBentryALTinterwordspacing
{AMPSE}. [Online]. Available: \url{https://github.com/USCPOSH/AMPSE}
\BIBentrySTDinterwordspacing

\bibitem{hassanpourghadi2019automated}
M.~Hassanpourghadi, Q.~Zhang, P.~Sharma, J.~Nam, S.~Su, S.~Chowdhury,
  J.~Sathyamoorthy, W.~Unglaub, F.~Wang, M.~Chen \emph{et~al.}, ``Automated
  analog mixed signal {IP} generator for {CMOS} technologies,''
  \emph{GOMACTech}, 2019.

\bibitem{kunal2019align}
K.~Kunal, M.~Madhusudan, A.~K. Sharma, W.~Xu, S.~M. Burns, R.~Harjani, J.~Hu,
  D.~A. Kirkpatrick, and S.~S. Sapatnekar, ``{ALIGN}: Open-source analog layout
  automation from the ground up,'' in \emph{Proceedings of the 56th Annual
  Design Automation Conference 2019}, 2019, pp. 1--4.

\bibitem{PTM_website}
\BIBentryALTinterwordspacing
{PTM}. [Online]. Available: \url{https://ptm.asu.edu}
\BIBentrySTDinterwordspacing

\end{thebibliography}
